\documentclass[preprint,fleqn]{elsarticle}
\usepackage{lineno,hyperref}

\usepackage{subfigure}

\journal{Journal of \LaTeX\ Templates}

\biboptions{sort&compress}

\begin{document}

\begin{frontmatter}

\title{Development of a 32-channel ASIC for an X-ray  APD Detector onboard the ISS}


\author[waseda-address,tokyotech-address]{Makoto Arimoto\corref{mycorrespondingauthor}}
\cortext[mycorrespondingauthor]{Corresponding author}
\ead{m.arimoto@aoni.waseda.jp, arimoto@hp.phys.titech.ac.jp}

\author[tokyotech-address]{Shohei Harita}
\author[tokyotech-address]{Satoshi Sugita}
\author[tokyotech-address]{Yoichi Yatsu}
\author[tokyotech-address]{Nobuyuki Kawai}
\author[ISAS-address]{Hirokazu Ikeda}
\author[ISAS-address]{Hiroshi Tomida}
\author[ISAS-address]{Naoki Isobe}
\author[JAXA-address]{Shiro Ueno}
\author[RIKEN-address]{Tatehiro Mihara}
\author[RIKEN-address]{Motoko Serino}
\author[TUS-address]{Takayoshi Kohmura}
\author[Aogaku-address]{Takanori Sakamoto}
\author[Aogaku-address]{Atsumasa Yoshida}
\author[Osaka-address]{Hiroshi Tsunemi}
\author[Wakasa-address]{Satoshi Hatori}
\author[Wakasa-address]{Kyo Kume}
\author[Wakasa-address]{Takashi Hasegawa}



\address[waseda-address]{Research Institute for Science and Engineering, Waseda University, 3-4-1, Ohkubo, Shinjuku, Tokyo, 169-8555, Japan}
\address[tokyotech-address]{Tokyo Institute of Technology, 2-12-1 Ookayama, Meguro-ku, Tokyo,  152-8550, Japan}
\address[ISAS-address]{Institute of Space and Astronautical Science (ISAS), 3-1-1 Yoshinodai, Sagamihara, Kanagawa,  229-8510, Japan}
\address[JAXA-address]{ISS Science Project Office, Institute of Space and Astronautical Science (ISAS), Japan Aerospace Exploration Agency (JAXA), 2-1-1 Sengen, Tsukuba, Ibaraki 305-8505, Japan}
\address[RIKEN-address]{MAXI team, RIKEN, 2-1 Hirosawa, Wako, Saitama 351-0198, Japan}
\address[TUS-address]{Department of Physics, Tokyo University of Science, 2641 Yamazaki, Noda, Chiba, 278-8510, Japan}
\address[Aogaku-address]{College of Science and Engineering, Department of Physics and Mathematics, Aoyama Gakuin University, 5-10-1 Fuchinobe, Chuo-ku, Sagamihara, Kanagawa 252-5258, Japan}
\address[Osaka-address]{Department of Earth and Space Science, Osaka University, 1-1 Machikaneyama-cho, Toyonaka, Osaka 560-0043, Japan}
\address[Wakasa-address]{The Wakasa Wan Energy Research Center, 64-52-1 Nagatani, Tsuruga, Fukui 914-0192, Japan}

\begin{abstract}
We report on the design and performance of a mixed-signal application specific integrated circuit (ASIC) dedicated to avalanche photodiodes (APDs) in order to detect hard X-ray emissions in a wide energy band onboard the International Space Station.
To realize wide-band detection from 20 keV to 1 MeV, we use Ce:GAGG scintillators, each coupled to an APD, with low-noise front-end electronics capable of achieving a minimum energy detection threshold of 20 keV.
 The developed ASIC has the ability to read out 32-channel APD signals 
 using 0.35 $\mu$m CMOS technology, and an analog amplifier at the input stage is designed to suppress the capacitive noise primarily arising from the large detector capacitance of the APDs. The ASIC achieves a performance of
 2099 $e^-$ + 1.5 $e^-$/pF at root mean square (RMS) with a wide 300 fC dynamic range.
 Coupling a reverse-type APD with a Ce:GAGG scintillator, we obtain an energy resolution of 
 6.7\% (FWHM) at 662 keV and a minimum detectable energy of 20 keV at room temperature (20 $^{\circ}$C).
  Furthermore, we examine the radiation tolerance for space applications by using a 90 MeV proton beam, confirming that the ASIC is free of single-event effects and can operate properly without serious degradation in analog and digital processing. 

\end{abstract}


\end{frontmatter}


\section{Introduction}
Monitoring hard X-ray emissions from high-energy astronomical objects such as gamma-ray bursts (GRBs), supernovae, active galactic nuclei, black holes, and neutron star binaries plays an important role in their identification. In particular,
observations in a {\it wide} energy band including hard X-rays are very crucial to identifying GRBs, 
because GRB  emission ranges between a few keV to a few MeV  \cite{1993ApJ...413..281B}.
 In addition, GRBs are one of the most reliable counterparts for detectable gravitational wave (GW) sources such as GW150914  \cite{2016PhRvL.116f1102A}. The detection of electromagnetic emission from GW sources would be critical for clarifying their origin by connecting them with known astronomical objects.

{\it WF-MAXI} \cite{2014SPIE.9144E..2PK} is a proposed mission on the International Space Station (ISS) to detect and localize X-ray transients, and is specifically designed to search the sky with a wide field of view (FoV).
{\it WF-MAXI} has two scientific instruments: the Soft X-ray Large Solid Angle Camera, which is sensitive to the 0.7--10 keV band, and the Hard X-ray Monitor (HXM), which is sensitive to the 20 keV--1 MeV band. 
Both instruments have a very wide FoV of $\sim$25\% of the entire sky.
In particular, HXM is responsible for monitoring parts of the sky that other observatories do not cover in the hard X-ray band.
HXM consists of a one-dimensional array of crystal scintillators (Ce-doped Gd$_3$Al$_2$Ga$_3$O$_{12}$; Ce:GAGG) coupled with avalanche photodiodes (APDs). By adopting an array structure combined with a coded mask or a passive shield around the scintillator array, HXM can localize X-ray transient sources to an accuracy of within a few degrees in one direction. If we use more than two sets of HXM with different angles, we can determine the two-dimensional source position. To realize this function, signals from each scintillator should be independently read out. As a consequence, analog front-end electronics are required for the HXM.

A primary objective of developing our application specific integrated circuit (ASIC) is to perform analog and digital signal processing on multiple APD signals, in order to measure energy spectra in a wide energy band between 20 keV--1 MeV.  Thus, a wide dynamic range is needed for the ASIC, and the analog amplifier of the ASIC must have sufficiently low noise to detect low-energy X-ray photons at an energy of  20 keV.
In addition, devices using a Complementary Metal Oxide Semiconductor (CMOS) in space applications, as in an in-orbit situation, radiation damage leads to serious issues, such as total dose effects and single event effects (SEEs). 
 To ensure the in-orbit performance of our ASIC, the radiation tolerance should be evaluated.

 Furthermore, to achieve a large detector effective area ($\sim$ 100 cm$^2$) using a 32-channel readout, each scintillator size should be $\sim$1$\times$1$\times$5 cm$^3$  with large APDs such as the S8664-55 (Hamamatsu: 0.5$\times$0.5 cm$^2$ APD area and 1.0$\times$1.0 cm$^2$ package area). In such a case, detector capacitance becomes $\sim$100 pF, where typically Si and CdTe detector capacitances are low and range from a few pF to 10 pF. Thus, to achieve our requirement, the series noise due to the large APD capacitance must be reduced when developing analog amplifiers in our ASIC.

In this paper, we report on a developed ASIC dedicated to APDs. 
Detailed specifications and the electrical design of the ASIC are described in Section \ref{Sec:Design} and 
the evaluated performance is shown in Section \ref{Sec:Performance}.  Briefly, our ASIC achieves a performance comparable to the high-performance IDEAS ASIC (VA140) with low noise (1400 $e^{-}$ at a 100-pF load)  and a wide dynamic range (200 fC) \cite{1674-1137-40-11-116101}. This creates a reasonable benchmark for ASICs with large detector capacitance for similar applications. 
The results of the ASICs' tolerance to proton irradiation are presented in Section \ref{Sec:RadiationTolerance}, and Section \ref{Sec:Conclusion} concludes the paper.

\section{Design of  TT02APD32}\label{Sec:Design}
\subsection{Overview of TT02APD32}\label{Sec:OverviewASIC}
The TT02APD32 ASIC (henceforth TT02) is designed to read out 32 APD signal channels. Its main function is to condition the signal with a good signal-to-noise ratio and feed them into a Wilkinson-type analog-to-digital converter (ADC).
The TT02 is fabricated based on the XFAB 0.35 $\mu$m CMOS technology with 4-metal and MIM (Metal-Insulator-Metal) options, and assembled in a ceramic package of 160 pins (QC-160360-WZ, KYOCERA).
The electrical architecture of the TT02 utilizes a property of the Open IP project, which has amassed technologies and the development of several radiation detectors (e.g., \cite{5756681}).
In particular, since the large APD capacitance of $\sim$100 pF leads to a corresponding large capacitive noise, the TT02 implements an ingenious noise reduction scheme (see Section \ref{Sec:InputFET}).

 SEEs occur stochastically, and one of the most serious SEEs is the single event latch-up (SEL).
In a CMOS structure, parasitic thyristors are formed due to the use of pMOS and nMOS transistors in a single silicon substrate.
 If a high-energy particle impacts the parasitic thyristor, the resulting large energy deposition generates many electrons and holes.
 Once the generated charge exceeds a critical limit in a certain region of the substrate, the thyristors activate, inducing an extremely large current into relevant circuits. Since its inherent positive feedback behavior (i.e., the large current lowers the threshold of the thyristor) causes thermal heating, it is never expected to automatically terminate until the power supply shuts down.
Thus, the occurrence of SEL may result in the serious and permanent functional failure of ASICs, and can ultimately prevent a successful scientific mission.
To improve the robustness of the TT02 against SEL, an epitaxial fabrication with p-type doping is used \cite{4493048}.

\begin{figure}
   \begin{center}
   \includegraphics[height=6cm]{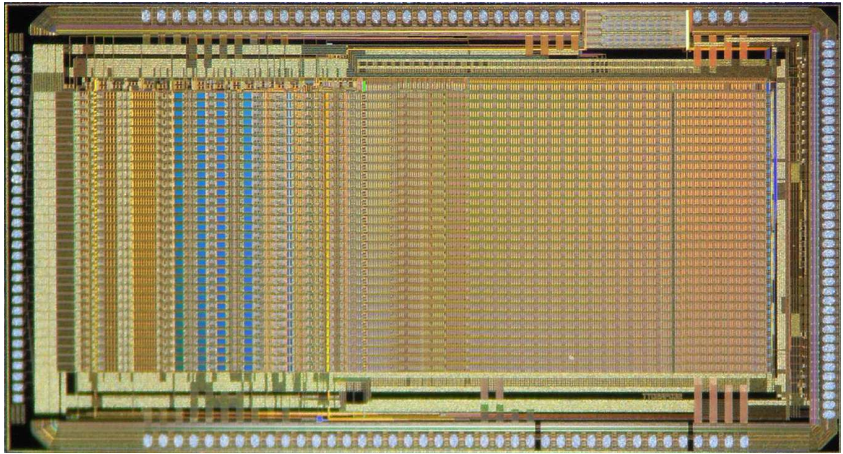}
   \end{center}
   \caption[example] 
   { \label{TT02APD32layout_label} 
Photograph of the developed ASIC TT02APD32. The dimensions are 8.4 $\times$ 4.8 mm$^2$.}
 \end{figure}

Figure \ref{TT02APD32layout_label} shows a photograph of the developed TT02 with a dimension of 4.8 $\times$ 8.4 mm$^2$. The specifications of the TT02 are shown in Table \ref{table1_label} and a schematic diagram of the TT02 is illustrated in Figure \ref{TT02APD32_Sch_label}.  

\subsection{Analog processing}\label{Sec:AnalogProcessing}
At an analog input, a charge sensitive amplifier (CSA) converts the input charge from the sensors or test pulse via $C_{\rm in}$ into a voltage through the feedback capacitance 
($C_{\rm f}$).
 The CSA output is directed into two paths: one, to a fast shaper with a peaking time of 0.5 $\mu$s, and the other to a slow
 shaper with a peaking time of 3 $\mu$s. Each signal channel is equipped with a pole-zero cancellation (PZC) circuit,
 which eliminates the long tail of the preamplifier output.
 The parameters for the PZC circuit are chosen to be 
 \begin{equation}
 R_{\rm f} C_{\rm f} =  R_{\rm pz} C_{\rm pz} =  R_{\rm pz}^\prime C_{\rm pz}^\prime.
 \end{equation}
 $R_{\rm f}$, $R_{\rm pz}$, and $R_{\rm pz}^\prime$ employ transfer-gate-type transistors with the application of an adjustable
 gate voltage $V_{\rm GG}$. Thus, $R_{\rm f}$ could be modified from a few 100 M$\Omega$ to $\sim$ G$\Omega$ by changing $V_{\rm GG}$. $R_{\rm pz}$ and $R_{\rm pz}^\prime$ are configured as a parallel combination of replicated $R_{\rm f}$ so as to maintain the appropriate PZC requirement.

 Since the PZC circuit does not include a time constant, the shaping amplifier is configured as a second-order 
low-pass filter, and the entire transfer function, including the PZC circuit, is equivalent to an ordinary CR--RC filter.
The peaking time of the slow shaper is determined by a set of $R_{\rm 1}$, $R_{\rm 2}$, $C_{\rm 1}$, and $C_{\rm 2}$. 
To obtain critical damping in order to reduce undershooting of the output signal, 
 the peaking time $\tau_{\rm peak, slow}$ = $C_{\rm 1}R_{\rm 1}$ = 4 $C_{\rm 2}R_{\rm 2}$ 
must be satisfied for the slow shaper. 
 Similarly for the fast shaper, we impose  
 $\tau_{\rm peak, fast}$ = $C_{\rm 1}^\prime R_{\rm 1}^\prime$ = 4 $C^\prime_{\rm 2}R_{\rm 2}^\prime$ on the peaking time.
 At the output stage of the slow shaper, the pulse height is acquired on a capacitor $C_{\rm H}$ at the timing of the HOLD signal, called the sample and hold (S/H). 

\begin{figure}
   \begin{center}
   \includegraphics[height=6cm]{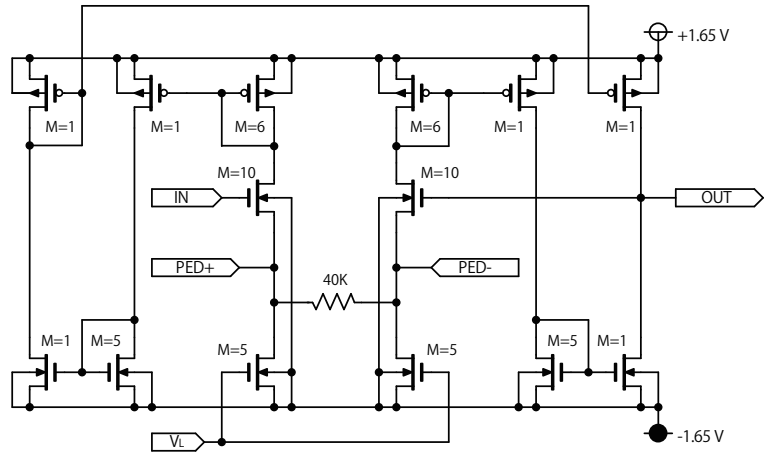}
   \end{center}
   \caption[example] 
   { \label{Current-mirror-register} 
 Current-mirror resistor employed as $R_{\rm 1}^\prime$ and $R_{\rm 2}^\prime$. A similar design is used for $R_{\rm 1}$ and $R_{\rm 2}$. 
By utilizing the asymmetric mirror ratio in a current-mirror pair and combining multiple pairs, a large resistor can be obtained. 
By feeding a constant current from PED+ to PED$-$, we can adjust the pedestal of the output signal. $V_{\rm L}$ is a constant voltage source.
}
 \end{figure}

 Here, we adopt a current-mirror resistor circuit ($R_{\rm 1}$, $R_{\rm 2}$, $R_{\rm 1}^\prime$ and $R_{\rm 2}^\prime$) to obtain a specified large resistance as illustrated in Figure \ref{Current-mirror-register}. 
The voltage difference between the IN/OUT nodes is converted into a
current by the poly resistor sandwiched by two nMOS transistors.
The circuit is configured to attenuate the current with the two stages of an asymmetric
 current mirror. As a result, the output current is substantially reduced to attain an
 effective resistance of a few M$\Omega$.
  Another benefit of this resistor circuit is its high-Z input impedance. As a consequence, the amplifier located just before the resistor circuit does not feed any current.
  In addition, by feeding an adjustable constant current into the resistor, a coarse pedestal adjustment can be done.

 Using a TSpice numerical simulation, we find that the stray capacitance around the current-mirror resistors distorts the waveform shape. To mitigate the pulse undershoot for the slow and fast shapers, the best choice for $C_{\rm 2}$ and $C_{\rm 2}^\prime$ is found to be a very low capacitance (a few fF) where we utilize nMOS transistors because their gate capacitance is very low, although this apparently violates the critical damping condition.

\subsection{Analog to digital conversion}
\begin{figure}
   \begin{center}
   \includegraphics[height=6cm]{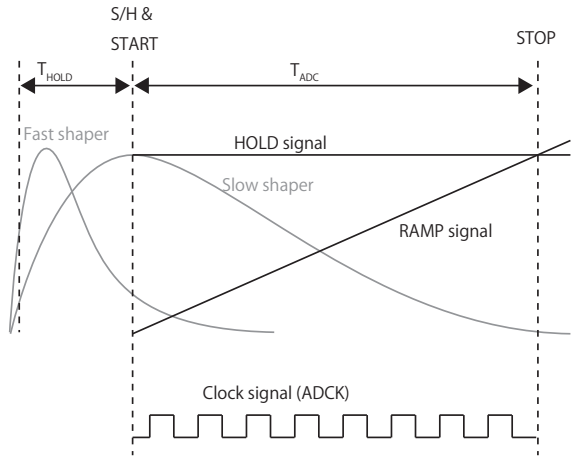}
   \end{center}
   \caption[example] 
   { \label{ADC_processing_procedure} 
Schematic view of the analog and digital processing for measuring the pulse height. Details are mentioned in the text.
  }
 \end{figure}

The output signal from the fast shaper goes into two comparators. One is used for the trigger signal, and to conduct the S/H at the slow shaper after the elapse of the $T_{\rm HOLD}$ period. 
  The other comparator detects very large signals coming from the APD, which is employed to quickly restore the baseline of the shaper output, and to subsequently mitigate the dead time. The discharge is conducted by the ``Release'' signal, which shorts the capacitor $C_{\rm 1}$ or $C_{\rm 1}^\prime$.
We utilize a Wilkinson-type ADC that converts the pulse height to the digital channel value by counting the number of clocks for the ADC (ADCK) from the input of the START signal until the input of the STOP signal (i.e., during the $T_{\rm ADC}$ period), where the STOP signal is generated when the RAMP signal exceeds the HOLD signal.
The schematic view of the analog and digital processing for the ADC function is illustrated in Figure \ref{ADC_processing_procedure}.
 The countable range for the ADC is 10 bits that are stored in digital registers.
 
 \subsection{Readout mode}
For each event, the measured ADC values are output by two readout modes: FORCED and SPARSE. While in the FORCED mode, all ADC values for all 32 channels are output, and in the SPARSE mode several ADC values are output only for selected channels that have a larger ADC value than a certain threshold. 
 Here, the SPARSE mode can be either analog or digital types, with the threshold levels determined by the analog voltage or the digital ADC value, respectively.
Furthermore, the TT02 detects a common-mode noise as described in detail in Section \ref{Sec:cmn-noise}. The ADC value of the measured common-mode noise can be read out for either mode.

By using the SPARSE mode, 
the amount of data is substantially reduced compared with the FORCED mode.  Thus, adoption of the SPARSE mode is important for space applications because in many in-orbit situations the amount of onboard memory is highly limited.

\begin{table}[h]
\caption{Specification of the TT02APD32 (TT02)}
\begin{center}
\begin{tabular}{lcc} \hline\hline\\[-6pt]
Number of channels        & 32  \\   
Dynamic range &  0 -- 300 fC\\
Peaking time &  \\
$\:\:\:\:$ fast shaper & 0.5 $\mu$s\\
$\:\:\:\:$ slow shaper &  3 $\mu$s\\
Integral non-linearity  &  2.5 \%\\
 Noise (RMS) at 100-pF load & 2211  $\pm$ 102 $e^{-}$\\
 Detectability of common-mode noise &  $>$1 ADC channel\\
Power supply & $\pm$1.65 V  \\
Power consumption    & $\sim$100 mW   \\ 
 SEU cross-section (90\% upper limit)    &  $<$2.2 $\times$ 10$^{-14}$ cm$^2$ bit$^{-1}$ \\ 
 Radiation tolerance    & $>$10.1 krad (101 Gy) \\   \hline 
\end{tabular}
\label{table1_label}
\end{center}
\end{table}

   \begin{figure}
   \begin{center}
   \includegraphics[height=6cm]{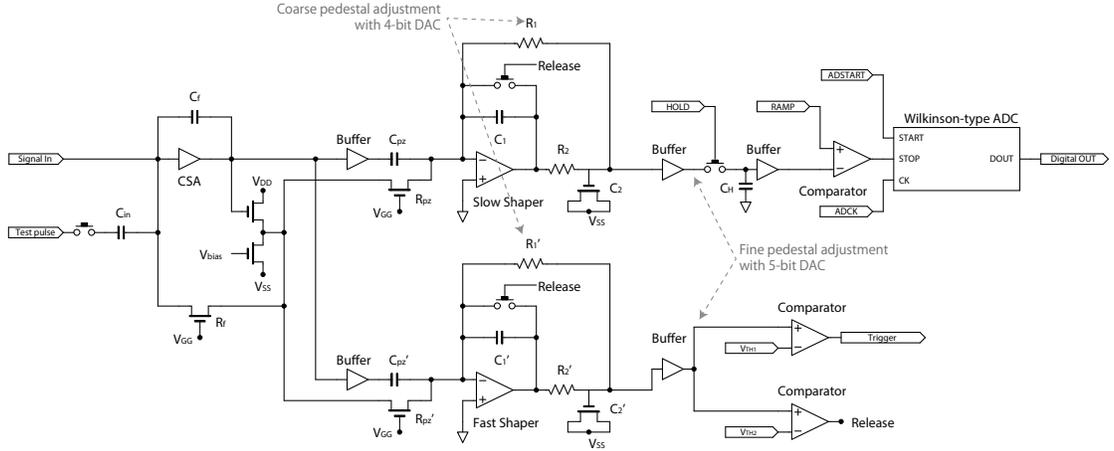}
   \end{center}
   \caption[example] 
   { \label{TT02APD32_Sch_label} 
 Schematic diagram of the signal processing for the TT02APD32.  
  }
 \end{figure}    

 \subsection{Input transistor of CSA}\label{Sec:InputFET}
\begin{figure}
   \begin{center}
   \includegraphics[height=6cm]{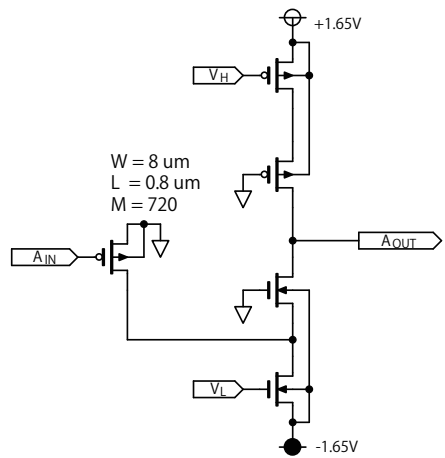}
   \end{center}
   \caption[example] 
   { \label{APD_preamp} 
Analog electrical circuit of the CSA dedicated to APDs. A folded-cascode amplifier is adopted, where the gate length, gate width, and number of transistors are denoted as $L$, $W$, and $M$. $V_{\rm H}$ and $V_{\rm L}$ are constant voltage sources. The APD's charge goes into $A_{\rm IN}$ and the amplified voltage signal is output from $A_{\rm OUT}$.
  }
 \end{figure}

The designed CSA is illustrated in Figure \ref{APD_preamp} where a folded-cascode amplifier is adopted to obtain a high gain and large output swing.
Since the APD's capacitance is large ($C_{\rm D}$ $\sim$ 100 pF), the capacitive noise is crucial for attaining a minimum detectable 
energy. To reduce the capacitive noise, the design of the input transistor of the CSA is key, and in the TT02, the noise power of the CSA is determined by the input transistor. 
The main contribution to capacitive noise power is a sum of the Johnson and flicker noise, which are proportional to $C_{\rm D}^2/g_{\rm m}\tau_{\rm peak}$ and $C_{\rm D}^2/A$, respectively, where $g_{\rm m}$ is the transconductance of the input transistor and $A$ is the gate area of the input transistor ($A$ = $WL$). 
$g_{\rm m}$ is a function of the drain current $I_{\rm D}$, the gate length $L$ and the gate width $W$.
 While in the strong inversion approximation $g_{\rm m}$ is proportional to  $\sqrt{I_{\rm D}W/L}$,
the actual operation point is near the weak inversion.
From the relation above, in order to reduce the capacitive noise power, larger $I_{\rm D}$, larger $W$, and moderate $L$ for larger $A$ are required. In addition, a pMOS transistor is well-known to have a smaller noise than that of a nMOS transistor and we use a pMOS transistor as an input transistor. 
Since a single transistor cannot have large gate area $A$ and $I_{\rm D}$, we use a number of pMOS transistors as an aggregated input transistor. 
Via numerical simulation using TSpice, we determined that $W$ = 8 $\mu$m and $L$ = 0.8 $\mu$m for the input transistor, and that in total 720 input transistors were required. The total $I_{\rm D}$, gate capacitance and $g_{\rm m}$ were estimated to be 793 $\mu$A, 18 pF, and 8.2 mS, respectively.

 \subsection{Dual Interlock Cell}\label{Sec:DICE}
\begin{figure}
   \begin{center}
   \includegraphics[height=6cm]{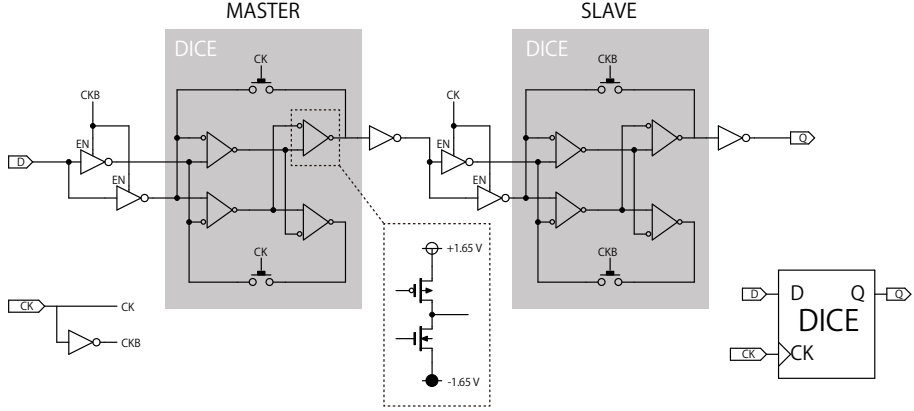}
   \end{center}
   \caption[example] 
   { \label{DICE_FF} 
Dual Interlock Cell (DICE) of the TT02 with a master-slave flip-flop scheme.
  }
 \end{figure}

 We adopt a master-slave flip-flop scheme to prevent malfunction due to distorted or noisy clock signals and stabilize the flip-flop's output. Furthermore, a countermeasure is needed for a single event upset (SEU), which affects the correct behavior of flip-flops.
If a high-energy particle impacts the CMOS circuits, the generated charge in the circuit accidentally activates the transistors, causing a bit inversion termed an SEU. Thus, an SEU event can significantly 
affect the ASIC operation when configured by control registers.
To prevent SEUs from occurring, Dual Interlock Cell (DICE) circuits are implemented in the digital circuit of the TT02, as illustrated in Figure \ref{DICE_FF}. By configuring a redundant dual structure, the original bit information is held even if one of the transistors in the DICE is accidentally activated. A similar DICE design has been widely used to protect ASICs from SEUs \cite{556880}.

\begin{figure}
   \begin{center}
   \includegraphics[height=6cm]{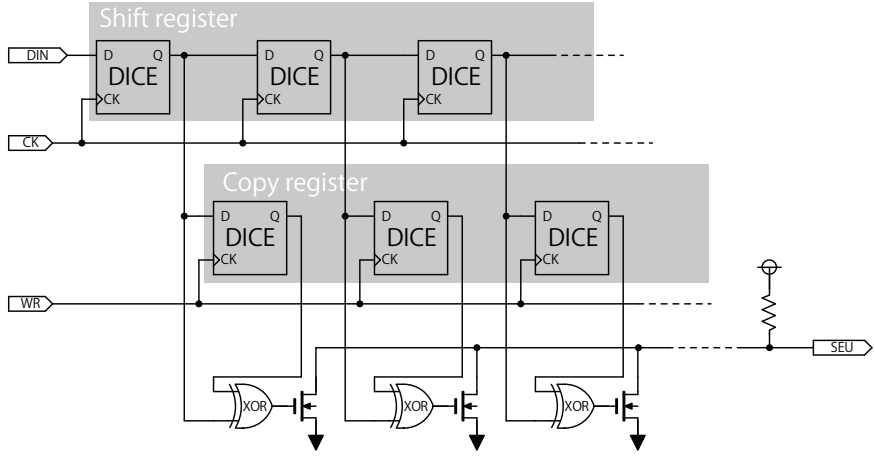}
   \end{center}
   \caption[example] 
   { \label{Shift_Register_DICE} 
Shift register implemented in the TT02 and capable of  detecting an SEU event. Digital bit data are duplicated on a copy register. If an SEU event occurs and the original bit is inverted in a certain DICE, an XOR element activates a transistor with a wired-OR connection.
  }
 \end{figure}

In addition, the TT02 can detect SEU events even if an SEU occurs in the TT02's digital registers.
 The digital registers holding the configuration settings for the TT02's signal processing are copied to separate registers, which are also DICE flip-flops as illustrated in Figure \ref{Shift_Register_DICE}. 
The differences between the content on the shift register and the copy are detected by XOR gates, to be summed up in a wired-OR scheme. In addition to the prompt output of the SEU alert signal, the SEU bit is stored into the readback data of the configuration register or the ADC data.
 
\subsection{Improvement from TT01}
The TT02 is a revision of a previous ASIC, TT01APD32 (henceforth TT01, \cite{2014SPIE.9144E..5ZA}). The TT01 has several problems and their related remedies in the TT02 are as follows:
\begin{itemize}
 \item  Tuning of the circuit parameters located around fast and slow shapers to suppress the degradation of the fast and slow waveforms (e.g., undershoot), as described in Section \ref{Sec:AnalogProcessing}.
 \item  Reduction of stray capacitances by analyzing the layout parasitic extraction and redesigning the schematic layout, alleviating distortion of the shapers' output waveform, as presented in Section \ref{Sec:Waveform}. 
 \item  Adding a new current source to independently adjust the constant voltages such as $V_{\rm L}$ in the current-mirror resistor and 4-bit DAC (digital to analog converter)  for coarse pedestal adjustment. Through this modification, we can tune the resistor values of $R_1$, $R_2$, $R_1^{\prime}$ and $R_2^{\prime}$ without any change of the pedestal DAC setting.
\end{itemize}

\section{Performance}\label{Sec:Performance}
\subsection{Waveform}\label{Sec:Waveform} 
\begin{figure}
\centering
\subfigure{\includegraphics[width=6.cm,angle=0]{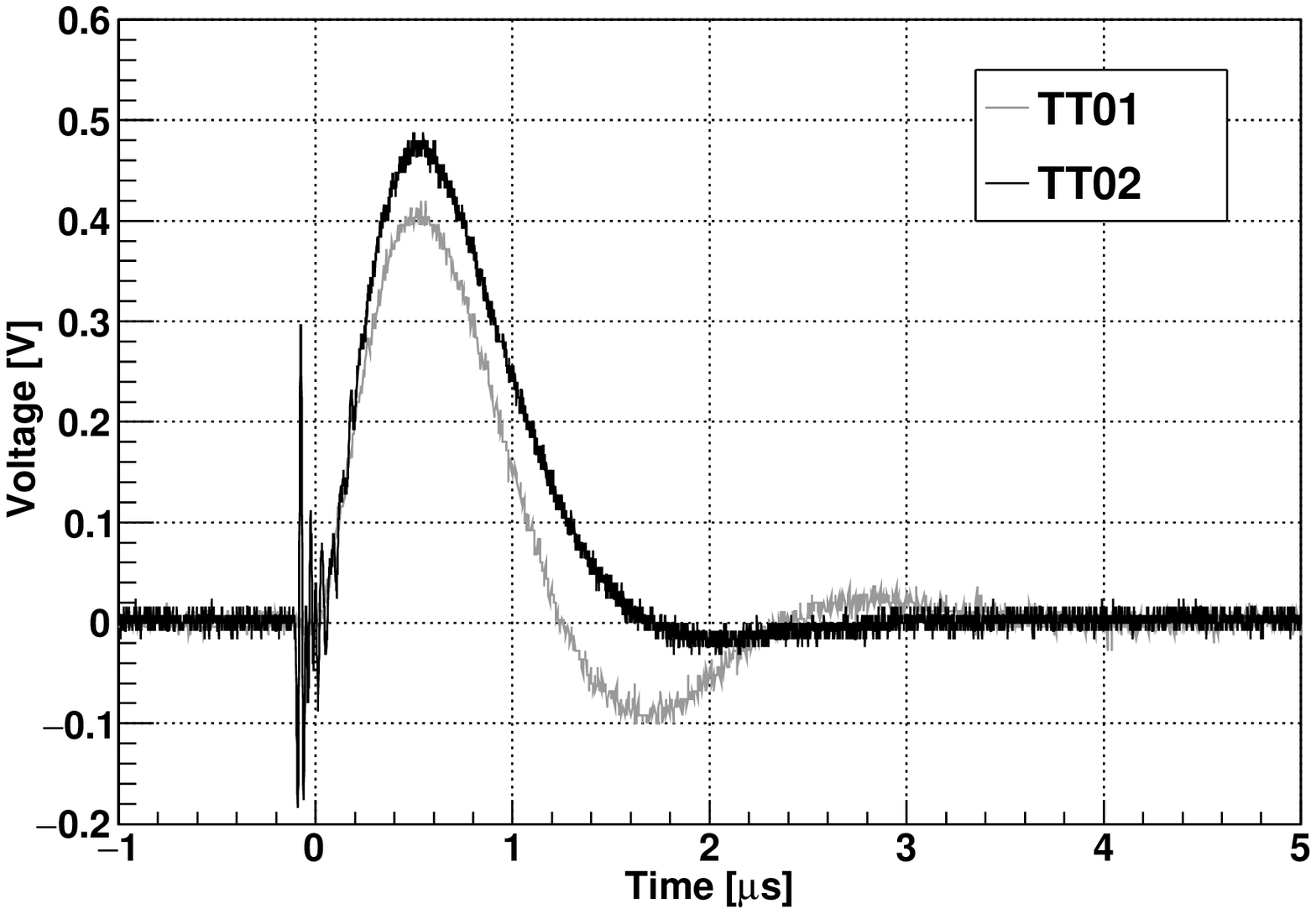}}
\subfigure{\includegraphics[width=6.cm,angle=0]{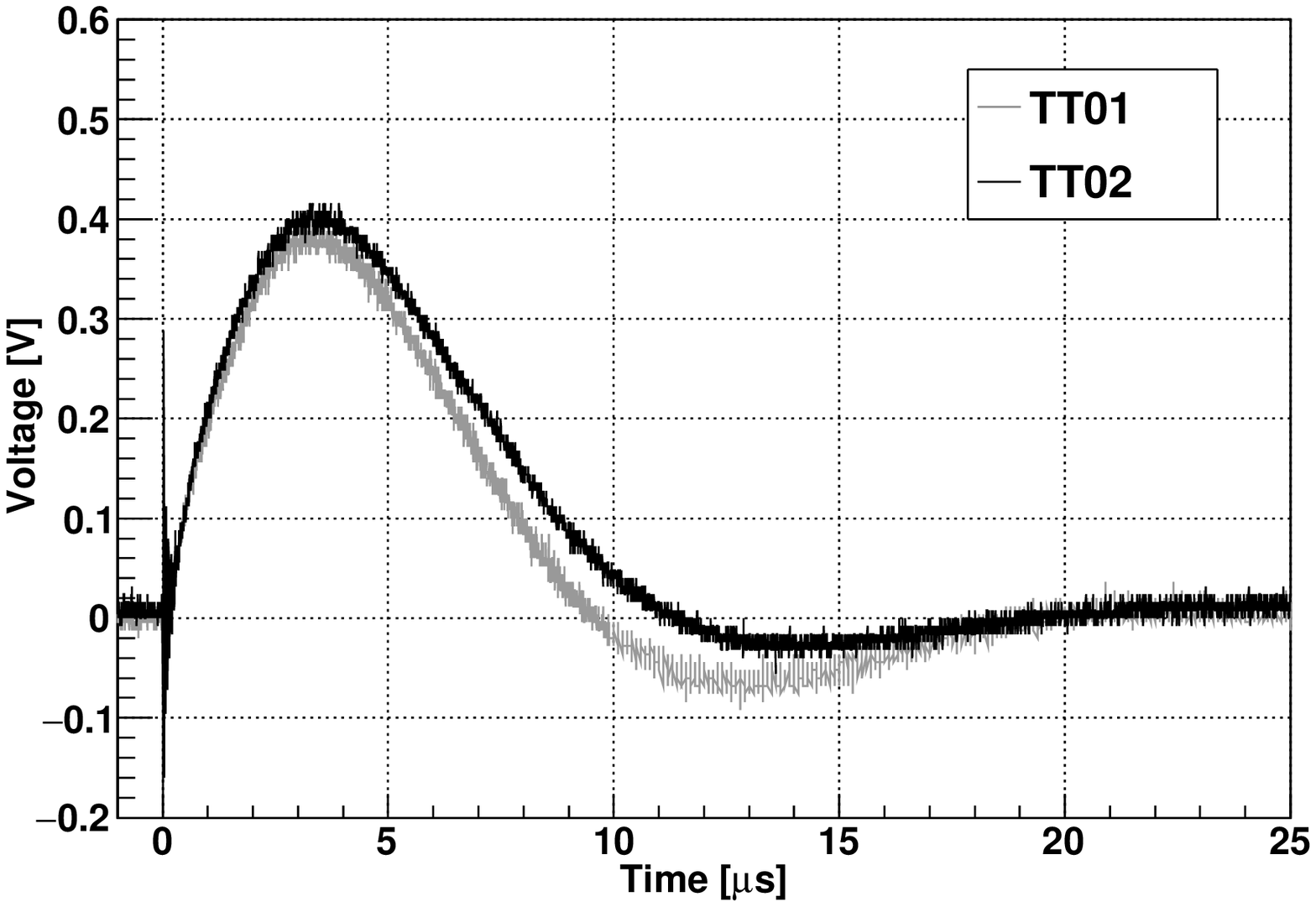}}
\caption{
{\it Left}: Waveforms of the fast shaper for the trigger. {\it Right}: Waveforms of the slow shaper for measuring the pulse height. TT01 (gray) means the previous version of the developed ASIC and TT02 (black) is the revised version including the reduction of stray capacitances located around the shapers, where the amount of the input charge is 150 fC.
}\label{fig:waveform_TT01_TT02}
\end{figure}

 For the TT01, a crucial problem is the degradation of the waveform of the fast and slow shapers, leading to lower pulse height, and more undershooting than expected, as shown in Figure \ref{fig:waveform_TT01_TT02}.
From our analysis of the layout parasitic extraction, we found that major stray capacitances are located amongst the fast/slow shapers and   power/ground planes in the TT01. Thus, we redesigned the electrical schematic layout to reduce the stray capacitances when developing the TT02.  

The improved TT02 waveforms are shown in Figure \ref{fig:waveform_TT01_TT02}. 
The undershoots and low pulse heights of the TT01 are mitigated. 
Additionally, the significant upgrading of the pulse height for the fast shaper by a factor of $\sim$20\% leads to an increase in the signal-to-noise ratio and a suppression of the noise level.

\subsection{Linearity of signal}
\begin{figure}
   \begin{center}
   \includegraphics[height=6cm]{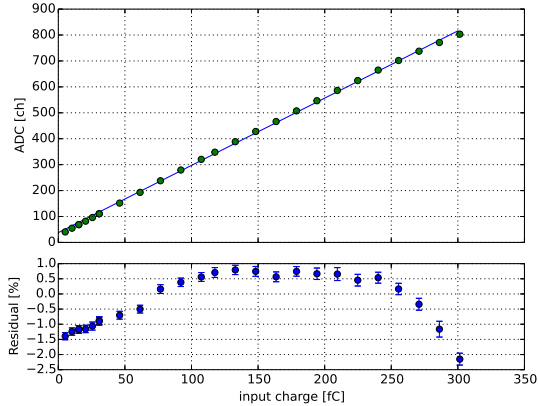}
   \end{center}
   \caption[example] 
   { \label{linearity_TT02} 
Linearity performance between the input charge and output ADC channel obtained with test pulses. The solid line denotes the best-fitting linear function. Residual is defined as (obtained value $-$ fitted value)/full scale.
}
 \end{figure}

 Here, we estimate the linearity performance between the input charge and output pulse.  Figure \ref{linearity_TT02} shows the linearity and its residual for a typical channel of the TT02, where the residual is defined as (obtained value $-$ fitted value)/full scale. The obtained linearity performance shows that the residuals range from  --2\% to  +0.5\%. 
 For small charge, the deviation from linear becomes larger because the fast shaper's waveform rises more slowly in that range and the trigger signal is delayed. For larger charge, the deviation also increases, but this is instead due to a saturation of the pulse height.
 The linearity performance is also characterized by integral non-linearity (INL). The INL for the TT02 is estimated to be 2.5 \%, which is comparable to other high-quality ASICs such as VA140 ($<$ 3\%) \cite{1674-1137-40-11-116101}.
 Furthermore, as described in Section \ref{Sec:Spectra}, the energy resolution, which is mainly determined by the light yield of the scintillator (e.g., 54.7 \% at 22 keV), is much larger than the systematic deviation due to the non-linearity. Thus, we conclude that the linearity performance of the TT02 is sufficient for spectroscopy.

\subsection{Detectability of common-mode noise}\label{Sec:cmn-noise}
 The TT02 can detect common-mode noise by obtaining a median of the 32-channel ADC values. To estimate the common-mode noise properly, the pedestal levels for every channel should be the same while the original pedestal levels for each channel have some variation. To adjust the pedestal levels, we employ coarse 4-bit current DACs implemented in the resistor circuits $R_{\rm 1}$ and $R_{\rm 1}^\prime$, and fine
 5-bit voltage DACs consisting of a resistor and a variable current source located at the back-end of the fast and slow shapers, as illustrated in Figure \ref{TT02APD32_Sch_label}.

\begin{figure}
   \begin{center}
   \subfigure{\includegraphics[width=5.cm,angle=0]{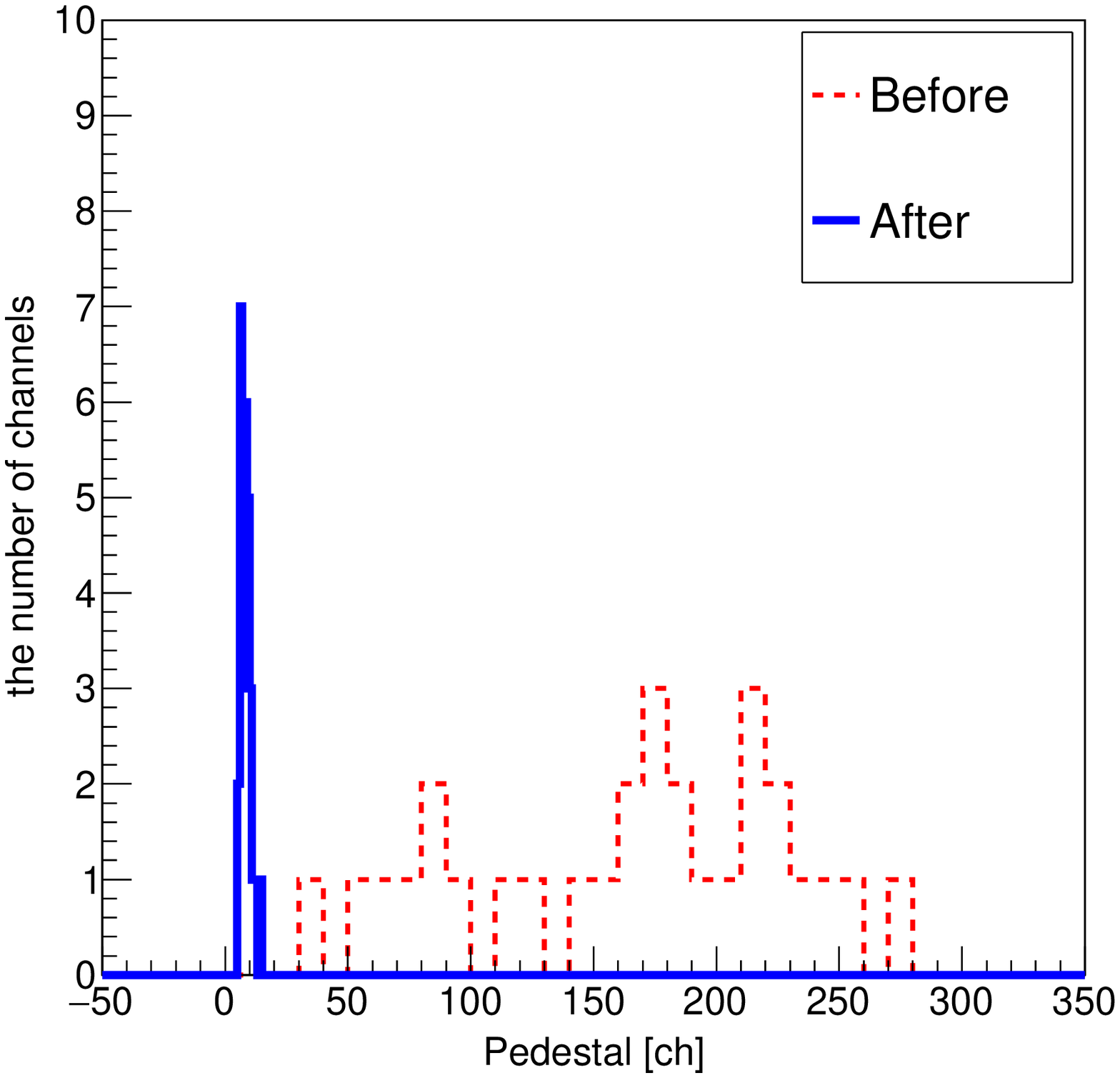}}
   \subfigure{\includegraphics[height=5.cm,angle=0]{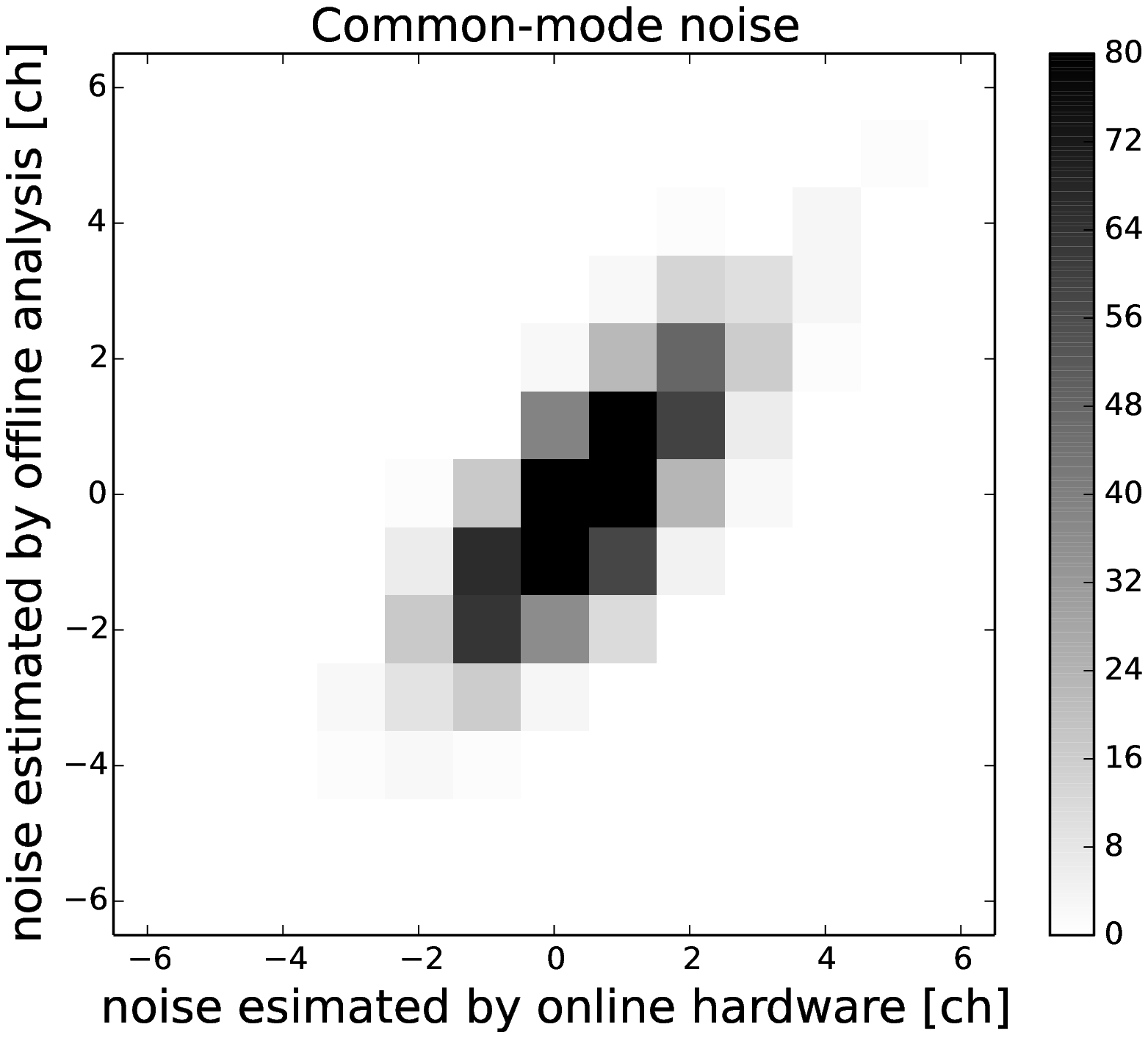}}
   \end{center}
   \caption[example] 
   { \label{cmn-noise} 
{\it Left}:  Pedestal distributions before and after the offset adjustment (before: 10-ch bin, after: 1-ch bin). The pedestal dispersions at root mean square are 64.1 ch and 2.1 ch, respectively.   {\it Right}: 
Common-mode noise distributions estimated by online hardware and offline analysis.
  }
 \end{figure}

  Figure \ref{cmn-noise} (left) shows the pedestal distribution, with and without adjustment. We find that the pedestal adjustment is well-executed within $\pm$2 ADC values. 
   Here, the original pedestal distribution has a large variation up to 300 ch which corresponds to a $\sim$ 200 mV offset voltage. Such a large offset voltage is mainly caused by the $R_{1}$ feedback transistors as shown in Figure \ref{TT02APD32_Sch_label}. 
 Due to manufacturing variations, the two transistors that sandwich the poly resistor have different threshold voltages $V_{\rm th}$, and several pairs of the current-mirror transistors are imbalanced. Additionally, the variation of $V_{\rm GS}$ of the CSA input transistor also contributes an offset voltage via $R_{\rm pz}$. These contributions conspire to create a large offset voltage, but a possible mitigation of the imbalance is to employ a common-centroid layout for the differential pair transistors and the current mirror.
  
We estimate the common-mode noise of test-pulse events in two ways. One is an {\it offline} procedure: in the FORCED readout mode, 32-channel ADC values are recorded for each event. By using all event data for every channel, the correction of the pedestal levels is very precisely executed by the software. By utilizing pedestal-corrected ADC values, we can estimate the {\it offline} common-mode noise.
The other approach is an {\it online} procedure, which is relatively simple compared to the offline analysis. Pedestal variations are corrected only by the hardware adjustment. The median value for each event is stored in digital registers of the TT02 without any correction by the software, and the stored median is the ``noise estimated by online hardware'' which is available in both FORCED and SPARSE modes.

The estimated common-mode noises with the offline and online ways are shown in Figure \ref{cmn-noise} (right), and we find that these common-mode noises are in close agreement. The linear correlation coefficient is 0.73, with a probability of $\sim$10$^{-165}$ that there is no correlation.
This result indicates that even in operation of the SPARSE mode, we can properly estimate the common-mode noise.
The offline method is more accurate in estimating the noise because the pedestal correction can be executed precisely. However, it needs 32-channel data in the FORCED mode, and a reasonable estimate of the common-mode noise in the SPARSE mode can reduce the volume of flight data by an order of magnitude, which would be beneficial for a bus system.

We note that hereafter, every ADC value is represented with common-mode-noise subtracted.
 In laboratory experiments such as noise measurement or spectrum acquisition using radioisotopes, common-mode noise is suppressed well and the subtraction of the common-mode noise does not significantly affect our result. However, in the proton beam test, the ASIC was operated in a noisy environment with a large common-mode noise and we used the detection of the common-mode noise as described in Section \ref{Sec:RadiationTolerance}.

\subsection{Equivalent Noise Charge}
\begin{figure}
   \begin{center}
   \includegraphics[height=6cm]{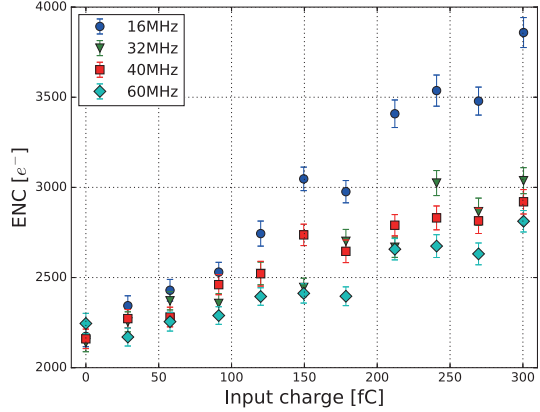}
   \end{center}
   \caption[example] 
   { \label{noise-frequency} 
Input charge versus equivalent noise charge (ENC) with different ADCK frequencies. The load capacitance is 100 pF.
In the higher frequency, a lower ENC is obtained.
  }
 \end{figure}

We plot measurements of the equivalent noise charge (ENC) for the slow shapers of the TT02 in Figure \ref{noise-frequency}. We find that the ENC value depends on the amount of input charge, which may be related to the ADC time period $T_{\rm ADC}$ 
because shortening it by increasing the ADCK frequency gives lower ENC values, where the gradient of the RAMP signal is adjusted to keep the same ADC value with different ADCK frequencies.
This might indicate that the $T_{\rm ADC}$-dependence is caused by an accumulation of external noise in the HOLD signal during the $T_{\rm ADC}$ period. 
To minimize this feature, we adopt a 60-MHz ADCK when obtaining energy spectra.

\begin{figure}
   \begin{center}
   \includegraphics[height=6cm]{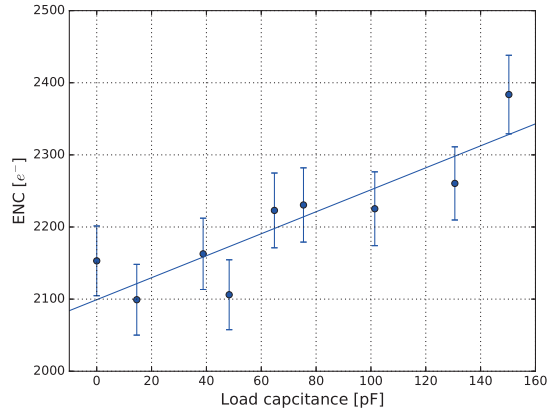}
   \end{center}
   \caption[example] 
   { \label{cap_gradient} ENC distribution with different load capacitances. The best fitting linear function (solid line) is
2099 $e^-$ + 1.5 $e^-$/pF. The measured data are taken from a channel which has a mean ENC as shown in Figure \ref{ENC_TT01vsTT02}.
  }
 \end{figure}

\begin{figure}
   \begin{center}
   \includegraphics[height=6cm]{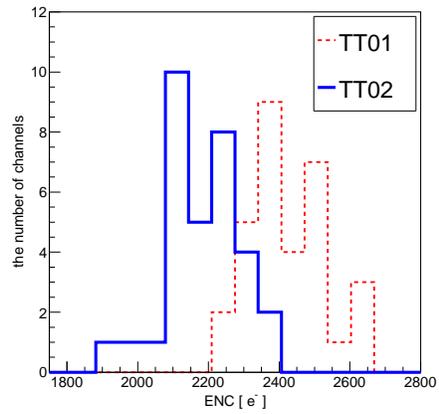}
   \end{center}
   \caption[example] 
   { \label{ENC_TT01vsTT02} 
Equivalent noise charge distributions in TT02 = 2211 $\pm$ 102 (RMS). TT01 = 2393 $\pm$ 107 (RMS). The load capacitance is 100 pF.
  }
 \end{figure}

To detect low-energy photons around a minimum detectable energy of 20 keV, the ENC of the TT02 in the small input charge range (i.e., $\sim$ 0 fC) should be small with a target value of $\sim$ 2000 $e^{-}$, considering APD dark current of 3 nA at room temperature (20 $^{\circ}$C). Figure \ref{cap_gradient} shows the measured ENC with different load capacitances at an input charge of 0 fC, where the 0-fC data can be obtained with a manual trigger mode without any charge input.
The obtained ENC around a typical APD's capacitance of 100 pF is $\sim$2200 $e^{-}$, which almost meets our requirement.
We show the ENC distributions of the TT01 and TT02 in Figure \ref{ENC_TT01vsTT02}. As mentioned in Section \ref{Sec:Waveform}, while for the TT01 the stray capacitances distort the fast and slow shapers' waveforms and degrade the signal-to-noise ratio, for the TT02 degradation of the waveforms is substantially reduced by redesigning the electrical layout, and we achieve a better ENC distribution in the TT02 by a factor of $\sim$10\% compared with that of the TT01.
 
  The obtained ENC is better than previous low-noise ASICs for APDs  such as 
 \cite{KOIZUMI2009327} ($\sim$3500 $e^-$ at a 100-pF load), and the wide dynamic range (300 fC) is a significant feature. Thus, the dynamic range to the noise ratio (D/N) is a good barometer for evaluating the ASIC performance quantitatively. We summarize these characteristics to compare previously reported high-quality ASICs for APDs or similar use in Table \ref{table2_label}, which indicates that the TT02 has performance comparable to the high-performance IDEAS ASIC (VA140).

\begin{table}[h]
\caption{{\bf Comparison of the ASIC performance}}
\begin{center}
\begin{tabular}{lccc} \hline\hline\\[-6pt]
        & ENC [$e^-$]$^\star$ & Dynamic range [$e^-$]  & D/N \\   \hline
 TT02 (this work) &   2200 &   1.87 $\times$ 10$^6$ &  850\\
 TIPPET08 \cite{KOIZUMI2009327}&  3460 &  2.85 $\times$ 10$^5$ &  80 \\
 CLEAR-PEM \cite{ALBUQUERQUE2009802}&  1300$^\dagger$ &   2.71 $\times$ 10$^5$ &  210 \\  
 VA140 \cite{1674-1137-40-11-116101} &  1400 &  1.25 $\times$ 10$^6$ &  890 \\  \hline 
\multicolumn{4}{l}{ $^\star$: 100-pF load}\\
\multicolumn{4}{l}{ $^\dagger$: ENC was reported only at a 10-pF load.}
\end{tabular}

\label{table2_label}
\end{center}
\end{table}

\subsection{Spectra}\label{Sec:Spectra}
To detect X-ray photons up to 1 MeV, we adopt Ce:GAGG scintillators with a high density of 6.63 g cm$^{-3}$. If we choose a 10 mm thickness, $\sim$20 \% of 1 MeV photons can be efficiently detected. 
Furthermore, a 520 nm peak wavelength of Ce:GAGG matches well with the APDs and a large light yield of 57,000 photons MeV$^{-1}$ leads to a good energy resolution and lowers the minimum detectable energy. In addition, Ce:GAGG does not have any hygroscopic effects which makes handling easier compared with other hygroscopic scintillators such as Tl:CsI or Tl:NaI.
 Here, we use a Hamamatsu reverse-type APD (S8664-55 with an APD area of 5 $\times$ 5 mm$^2$) which is widely used as a scintillation detector \cite{2003NIMPA.515..671I}, with its space use verified in small satellite projects \cite{2010JGRA..115.5204K,2014SPIE.9144E..0LY}.

\begin{figure}
   \begin{center}
   \includegraphics[height=6cm]{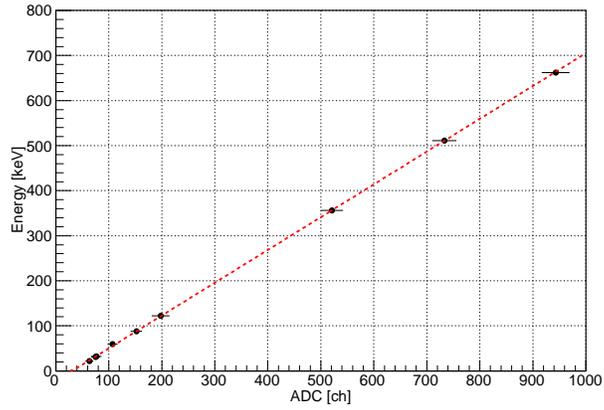}
   \end{center}
   \caption[example] 
   { \label{ch_vs_energy} 
Plot between ADC channel versus energy. The data points were obtained by measuring several radioactive sources such as $^{109}$Cd, $^{241}$Am, $^{133}$Ba, $^{57}$Co, $^{22}$Na and $^{137}$Cs. The dashed line represents the best-fit linear function.
  }
 \end{figure}

\begin{figure}
   \begin{center}
   \includegraphics[height=6cm]{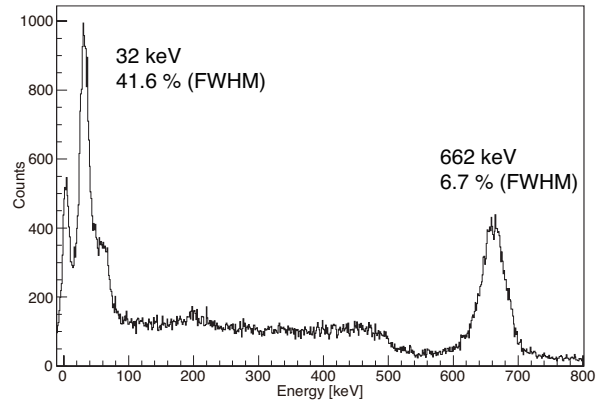}
   \end{center}
   \caption[example] 
   { \label{spec_137Cs} 
Energy spectra of $^{137}$Cs measured with Ce:GAGG + a reverse-type APD (S8664-55) + TT02, measured at 20 $^{\circ}$C. Obtained energy resolutions are 41.6\% (FWHM) at 32 keV and 6.7\% (FWHM) at 662 keV. 
  }
 \end{figure}

\begin{figure}
   \begin{center}
   \includegraphics[height=6cm]{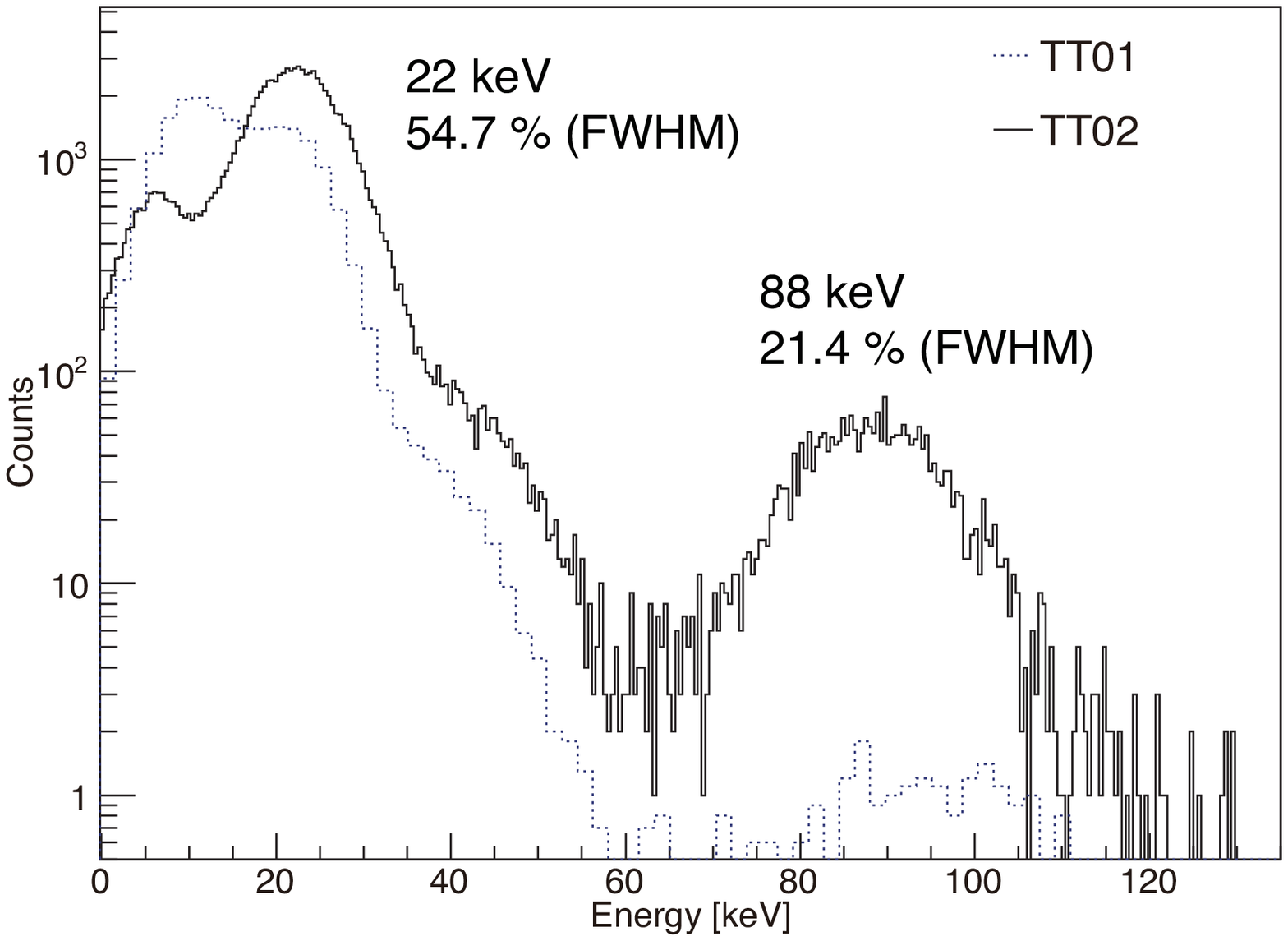}
   \end{center}
   \caption[example] 
   { \label{spec_109Cd} 
   Energy spectra of $^{109}$Cd measured with Ce:GAGG + a reverse-type APD (S8664-55) + TT01 (dashed line) and TT02 (solid line), measured at 20 $^{\circ}$C. Obtained energy resolutions measured with TT02 are 54.7\% (FWHM) at 22 keV and 21.4\% (FWHM) at 88 keV. Note that the counts of TT01 are scaled by 0.1.  Here, in the TT01 spectrum the loss of 88 keV is due to a high noise rate and significant increase of a dead time.
  }
 \end{figure}

  By using a Ce:GAGG scintillator of 10 $\times$ 10 $\times$ 50 mm$^3$, a reverse-type APD of 5 $\times$ 5 mm$^2$ and the TT02, we obtained spectra from several radioactive sources such as $^{109}$Cd, $^{241}$Am, $^{133}$Ba, $^{57}$Co, $^{22}$Na and $^{137}$Cs, measured at 20 $^{\circ}$C and an avalanche gain of the APD at 100.  The ADC channel and X-ray energy are linearly related, as shown in Figure \ref{ch_vs_energy}. Here we show the representative spectra of $^{137}$Cs  (Figure \ref{spec_137Cs}) and $^{109}$Cd (Figure \ref{spec_109Cd}).
For the irradiation of $^{137}$Cs, a 32 keV line with an energy resolution of 41.6\% (full width at half maximum: FWHM) is clearly detected, and the energy resolution at 662 keV is 6.7\% (FWHM). To estimate the minimum detectable energy, we use the 22 keV line of $^{109}$Cd. Figure \ref{spec_109Cd} shows the $^{109}$Cd spectra obtained with the TT01 and TT02. While for the TT01 the 22-keV line is barely detected, for the TT02 it is clearly detected due to the improvement of the waveforms for fast and slow shapers with a larger signal-to-noise ratio. In particular,
the improvement of the fast shaper's waveform makes the contribution from the noise around 10 keV lower. Thus, we achieve the minimum detectable energy of 20 keV.

 \section{Radiation tolerance}\label{Sec:RadiationTolerance}
\begin{figure}
   \begin{center}
   \includegraphics[height=6cm]{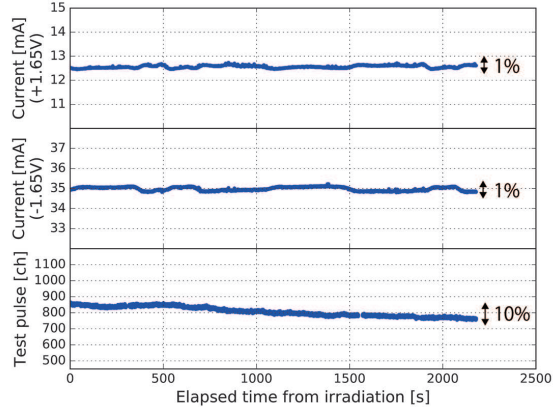}
   \end{center}
   \caption[example] 
   { \label{tp_variation} 
Temporal variation of the analog currents for +1.65 V/--1.65 V and the ADC value of test-pulse event.
  }
 \end{figure}

 To evaluate the radiation tolerance of the TT02, a proton irradiation test at 90 MeV was performed at the Wakasa Wan Energy Research Center in Fukui, Japan. We used 90 MeV protons because the cross section of the SEE for digital electrical elements rapidly increases from $\sim$10 MeV and becomes saturated around 100 MeV \cite{736527, 1546461}.  The proton beam is directly irradiated on the ASIC without any cover or cap, and every proton penetrates the ASIC because the proton range is $\sim$ 3.5 cm in the Si medium, which is much longer than the thickness of the ASIC ($<$1 mm).
The beam rate was 4.8 $\times$ 10$^{7}$ protons cm$^{-2}$ s$^{-1}$ with a duration of 2160 s. The total number of irradiated protons was 1.0 $\times$ 10$^{11}$ protons cm$^{-2}$, which corresponds to protons accumulated during 160 years in ISS orbit, where the in-orbit proton flux is calculated by the SPENVIS tool \cite{Spenvis}.
 During the proton beam irradiation, we monitored the analog currents of +1.65 V and $-$1.65 V, as shown in Figure \ref{tp_variation}. If SEL events were to occur in the TT02, the monitored currents would be expected to drastically increase.
The result shows that the temporal variations of the analog currents are $\sim$1 \% and there is no sign of SEL events.
In addition, after monitoring the SEU bit as illustrated in Figure \ref{Shift_Register_DICE}, we did not obtain any SEU events during the proton irradiation. Considering that the TT02 has 1030-bit registers, the SEU cross-section of the TT02 is estimated to be $\sigma_{\rm SEU}$ $<$ 2.2 $\times$ 10$^{-14}$ cm$^2$ bit$^{-1}$ (90\% upper limit).

  In the total dose effect, accumulation of charge occurs in the gate oxide, and continuing damage causes an increase of the leakage current and a shift in the voltage threshold of the transistors.
The total dose of the proton irradiation is 10.1 krad (101 Gy), which corresponds to an in-orbit dose for $\sim$10 years because the typical dose rate in a low-earth orbit is roughly 1 krad year$^{-1}$ (10 Gy year$^{-1}$).
During the proton irradiation, test-pulse signals were continuously injected into the TT02. The ADC values of the test pulse gradually decreased by a factor of 10 \%, while there was no significant difference between the measured pulse widths of the test-pulse events before and after the proton irradiation (1.11$\pm$0.05 ch versus 1.04$\pm$0.03 ch
  after the subtraction of the common-mode noise). 
 Here, the measured pulse widths were affected by external noise, and before the subtraction of the common-mode noise we obtained original pulse widths of $\sim$ 2.2 ch. Thus, the detection of the common-mode noise as mentioned in Section \ref{Sec:cmn-noise} is useful for estimating and subtracting the external noise. 

 As shown in Figure \ref{tp_variation} although the 10\% decrease of the pulse height after 10 years from launch slightly degrades the performance of the TT02, measured for example by an increase in the minimum detectable energy, this result indicates that the total dose effect does not have a serious impact on the main functions of the TT02, considering that a typical mission life is $<$ 10 years.
Thus, we conclude that our developed ASIC has sufficient radiation tolerance for decadal-scale space applications, when protons are considered.

 \section{Conclusion}\label{Sec:Conclusion}
 We have developed an analog and digital ASIC (TT02) dedicated to the readout of APDs combined with Ce:GAGG scintillators.
By suppressing stray capacitances, the TT02 achieves a low-noise performance  ($\sim$2200 $e^-$) for a large detector capacitance  of 100 pF with a wide dynamic range of a 300 fC, and applications to spectroscopy from 20 keV to 1 MeV can be realized.
 The TT02 provides an excellent ability to detect common-mode noise with a sensitivity of $>$ 1 ADC channel, which also enables a substantial reduction of the total amount of mission data by an order of magnitude. 
  Through proton irradiation testing, we have demonstrated that  the TT02 has a sufficient radiation tolerance for decadal-scale space use, and (1) is free of any SEE events such as SEL and SEU with $\sigma_{\rm SEU}$ $<$ 2.2 $\times$ 10$^{-14}$ cm$^2$ bit$^{-1}$ (90\% upper limit), and (2) the total dose effect provides only a weak degradation  (10\%) for the spectroscopic applications of the TT02, while the primary functions work properly.

\section*{Acknowledgement}
This work was supported by JSPS KAKENHI Grant Number JP24103002,  JP23740199, JP18740110 and the Key Researchers Development Program at Waseda University.


\bibliography{mybibfile}

\begin{thebibliography}{10}
\expandafter\ifx\csname url\endcsname\relax
  \def\url#1{\texttt{#1}}\fi
\expandafter\ifx\csname urlprefix\endcsname\relax\def\urlprefix{URL }\fi
\expandafter\ifx\csname href\endcsname\relax
  \def\href#1#2{#2} \def\path#1{#1}\fi

\bibitem{1993ApJ...413..281B}
D.~{Band}, J.~{Matteson}, L.~{Ford}, B.~{Schaefer}, D.~{Palmer},
  B.~{Teegarden}, T.~{Cline}, M.~{Briggs}, W.~{Paciesas}, G.~{Pendleton},
  G.~{Fishman}, C.~{Kouveliotou}, C.~{Meegan}, R.~{Wilson}, P.~{Lestrade},
  {BATSE observations of gamma-ray burst spectra. I - Spectral diversity},
  Astrophysical Journal 413 (1993) 281--292.
\newblock \href {http://dx.doi.org/10.1086/172995} {\path{doi:10.1086/172995}}.

\bibitem{2016PhRvL.116f1102A}
B.~P. {Abbott}, R.~{Abbott}, T.~D. {Abbott}, M.~R. {Abernathy}, F.~{Acernese},
  K.~{Ackley}, C.~{Adams}, T.~{Adams}, P.~{Addesso}, R.~X. {Adhikari}, et~al.,
  {Observation of Gravitational Waves from a Binary Black Hole Merger},
  Physical Review Letters 116~(6) (2016) 061102.
\newblock \href {http://arxiv.org/abs/1602.03837} {\path{arXiv:1602.03837}},
  \href {http://dx.doi.org/10.1103/PhysRevLett.116.061102}
  {\path{doi:10.1103/PhysRevLett.116.061102}}.

\bibitem{2014SPIE.9144E..2PK}
N.~{Kawai}, H.~{Tomida}, Y.~{Yatsu}, T.~{Mihara}, S.~{Ueno}, M.~{Kimura},
  M.~{Arimoto}, M.~{Serino}, T.~{Sakamoto}, H.~{Tsunemi}, T.~{Kohmura},
  H.~{Negoro}, Y.~{Ueda}, M.~{Morii}, Y.~{Tsuboi}, K.~{Ebisawa}, A.~{Yoshida},
  {Wide-field MAXI: soft x-ray transient monitor on the ISS}, in: Space
  Telescopes and Instrumentation 2014: Ultraviolet to Gamma Ray, Vol. 9144 of
  Proceedings of the SPIE, 2014, p. 91442P.
\newblock \href {http://dx.doi.org/10.1117/12.2057188}
  {\path{doi:10.1117/12.2057188}}.

\bibitem{1674-1137-40-11-116101}
F.~Zhang, W.-X. Peng, K.~Gong, D.~Wu, Y.-F. Dong, R.~Qiao, R.-R. Fan, J.-Z.
  Wang, H.-Y. Wang, X.~Wu, D.~L. Marra, P.~Azzarello, V.~Gallo, G.~Ambrosi,
  A.~Nardinocchi,
  \href{http://stacks.iop.org/1674-1137/40/i=11/a=116101}{Design of the readout
  electronics for the dampe silicon tracker detector}, Chinese Physics C
  40~(11) (2016) 116101.
\newline\urlprefix\url{http://stacks.iop.org/1674-1137/40/i=11/a=116101}

\bibitem{5756681}
G.~Sato, T.~Kishishita, H.~Ikeda, T.~Sakumura, T.~Takahashi, Development of
  low-noise front-end asic for hybrid cdte pixel detectors, IEEE Transactions
  on Nuclear Science 58~(3) (2011) 1370--1375.
\newblock \href {http://dx.doi.org/10.1109/TNS.2011.2135865}
  {\path{doi:10.1109/TNS.2011.2135865}}.

\bibitem{4493048}
G.~F.~E. Gene, N.~C. Lee, T.~K. Tong, D.~Sim, Impact on latchup immunity due to
  the switch from epitaxial to bulk substrate, in: 2006 IEEE International
  Symposium on Semiconductor Manufacturing, 2006, pp. 156--159.
\newblock \href {http://dx.doi.org/10.1109/ISSM.2006.4493048}
  {\path{doi:10.1109/ISSM.2006.4493048}}.

\bibitem{556880}
T.~Calin, M.~Nicolaidis, R.~Velazco, Upset hardened memory design for submicron
  cmos technology, IEEE Transactions on Nuclear Science 43~(6) (1996)
  2874--2878.
\newblock \href {http://dx.doi.org/10.1109/23.556880}
  {\path{doi:10.1109/23.556880}}.

\bibitem{2014SPIE.9144E..5ZA}
M.~{Arimoto}, Y.~{Yatsu}, N.~{Kawai}, H.~{Ikeda}, A.~{Harayama}, S.~{Takeda},
  T.~{Takahashi}, H.~{Tomida}, S.~{Ueno}, M.~{Kimura}, T.~{Mihara},
  M.~{Serino}, H.~{Tsunemi}, A.~{Yoshida}, T.~{Sakamoto}, T.~{Kohmura},
  H.~{Negoro}, Y.~{Ueda}, {Development of the hard x-ray monitor onboard
  WF-MAXI}, in: Space Telescopes and Instrumentation 2014: Ultraviolet to Gamma
  Ray, Vol. 9144 of Proceedings of the SPIE, 2014, p. 91445Z.
\newblock \href {http://dx.doi.org/10.1117/12.2054942}
  {\path{doi:10.1117/12.2054942}}.

\bibitem{KOIZUMI2009327}
M.~Koizumi, J.~Kataoka, S.~Tanaka, H.~Ishibashi, N.~Kawai, H.~Ikeda,
  Y.~Ishikawa, N.~Kawabata, Y.~Matsunaga, K.~Shimizu, H.~Kubo, Development of a
  low-noise analog front-end asic for apd-pet detectors, Nuclear Instruments
  and Methods in Physics Research Section A: Accelerators, Spectrometers,
  Detectors and Associated Equipment 604~(1) (2009) 327 -- 330, pSD8.
\newblock \href
  {http://dx.doi.org/http://dx.doi.org/10.1016/j.nima.2009.01.104}
  {\path{doi:http://dx.doi.org/10.1016/j.nima.2009.01.104}}.

\bibitem{ALBUQUERQUE2009802}
E.~Albuquerque, V.~Bexiga, R.~Bugalho, B.~Carriço, C.~S. Ferreira,
  M.~Ferreira, J.~Godinho, F.~Gonçalves, C.~Leong, P.~Lousã, P.~Machado,
  R.~Moura, P.~Neves, C.~Ortigão, F.~Piedade, J.~F. Pinheiro, J.~Rego,
  A.~Rivetti, P.~Rodrigues, J.~C. Silva, M.~M. Silva, I.~C. Teixeira, J.~P.
  Teixeira, A.~Trindade, J.~Varela, Experimental characterization of the 192
  channel clear-pem frontend asic coupled to a multi-pixel apd readout of
  lyso:ce crystals, Nuclear Instruments and Methods in Physics Research Section
  A: Accelerators, Spectrometers, Detectors and Associated Equipment 598~(3)
  (2009) 802 -- 814.
\newblock \href
  {http://dx.doi.org/http://dx.doi.org/10.1016/j.nima.2008.10.005}
  {\path{doi:http://dx.doi.org/10.1016/j.nima.2008.10.005}}.

\bibitem{2003NIMPA.515..671I}
T.~{Ikagawa}, J.~{Kataoka}, Y.~{Yatsu}, N.~{Kawai}, K.~{Mori}, T.~{Kamae},
  H.~{Tajima}, T.~{Mizuno}, Y.~{Fukazawa}, Y.~{Ishikawa}, N.~{Kawabata},
  T.~{Inutsuka}, {Performance of large-area avalanche photodiode for low-energy
  X-rays and {$\gamma$}-rays scintillation detection}, Nuclear Instruments and
  Methods in Physics Research A 515 (2003) 671--679.
\newblock \href {http://dx.doi.org/10.1016/j.nima.2003.07.024}
  {\path{doi:10.1016/j.nima.2003.07.024}}.

\bibitem{2010JGRA..115.5204K}
J.~{Kataoka}, T.~{Toizumi}, T.~{Nakamori}, Y.~{Yatsu}, Y.~{Tsubuku},
  Y.~{Kuramoto}, T.~{Enomoto}, R.~{Usui}, N.~{Kawai}, H.~{Ashida},
  K.~{Omagari}, K.~{Fujihashi}, S.~{Inagawa}, Y.~{Miura}, Y.~{Konda},
  N.~{Miyashita}, S.~{Matsunaga}, Y.~{Ishikawa}, Y.~{Matsunaga}, N.~{Kawabata},
  {In-orbit performance of avalanche photodiode as radiation detector on board
  the picosatellite Cute-1.7+APD II}, Journal of Geophysical Research (Space
  Physics) 115 (2010) A05204.
\newblock \href {http://dx.doi.org/10.1029/2009JA014699}
  {\path{doi:10.1029/2009JA014699}}.

\bibitem{2014SPIE.9144E..0LY}
Y.~{Yatsu}, K.~{Ito}, S.~{Kurita}, M.~{Arimoto}, N.~{Kawai}, M.~{Matsushita},
  S.~{Kawajiri}, S.~{Kitamura}, S.~{Matunaga}, S.~{Kimura}, J.~{Kataoka},
  T.~{Nakamori}, S.~{Kubo}, {Pre-flight performance of a micro-satellite
  TSUBAME for X-ray polarimetry of gamma-ray bursts}, in: Space Telescopes and
  Instrumentation 2014: Ultraviolet to Gamma Ray, Vol. 9144 of Proceedings of
  the SPIE, 2014, p. 91440L.
\newblock \href {http://dx.doi.org/10.1117/12.2056275}
  {\path{doi:10.1117/12.2056275}}.

\bibitem{736527}
T.~Goka, H.~Matsumoto, N.~Nemoto, See flight data from japanese satellites,
  IEEE Transactions on Nuclear Science 45~(6) (1998) 2771--2778.
\newblock \href {http://dx.doi.org/10.1109/23.736527}
  {\path{doi:10.1109/23.736527}}.

\bibitem{1546461}
Y.~Kimoto, N.~Nemoto, H.~Matsumoto, K.~Ueno, T.~Goka, T.~Omodaka, Space
  radiation environment and its effects on satellites: analysis of the first
  data from teda on board adeos-ii, IEEE Transactions on Nuclear Science 52~(5)
  (2005) 1574--1578.
\newblock \href {http://dx.doi.org/10.1109/TNS.2005.855822}
  {\path{doi:10.1109/TNS.2005.855822}}.

\bibitem{Spenvis}
http://www.spenvis.oma.be/.

\end{thebibliography}

\end{document}